# J-measure of uncertainty (JMU)
# for specified probability distribution

Y. Schreiber*, A. Chudnovsky**

 *  Kurt-Schumacher-Str.11, 86165 Augsburg, Germany
** The University of Illinois At Chicago, Chicago IL, 60607, USA

## Abstract.

In this paper it is shown that in the case of the given probability distribution or histogram using the Jaynes entropy maximum principle an J- measure of uncertainty (JMU) provides the maximum for a given distribution can be constructed. Formulas for the introduced JMU were obtained explicitly and calculations of this new measure for a number of distributions are shown as examples. It is shown using as the example a two-dimensional random variable, the application of the proposed method to the JMU estimation for the multidimensional case. It was made a comparison of the information contained in the histogram of a random variable with the information in the probability distribution obtained as a fitting of this histogram. Moreover  studied the influence of an additional measurement of a certain physical quantity on the amount of information.

## 1. Introduction.

This paper originated as the intersection of two ideas:  generalized entropies and Jaynes maximum entropy principle. However, the named after Jaynes J-measure of uncertainty, proposed in this paper, is not a generalization of the classical Boltzmann-Gibbs (BG) entropy. In essence, the only starting point for the proposed method is maximum entropy principle.

A. Ya. Khinchin formulated four axioms, which are necessary and sufficient conditions for the entropy to have the form of Gibbs entropy [1]. It is shown in [2-4] that if one of these axioms fails, namely the axiom of additivity, a two-parameter family of entropies of a certain structure appears. Moreover, in this and all other generalizations of BG entropy, the latter is the simplest particular case.

However, it should be noted that Khinchin's axioms were formulated by him for discrete random variables. For continuous random variables BG entropy is used rather by analogy. In this case, some problems arise, so that Kullback-Leibler divergence, also called relative entropy, is used more often. For this reason, in particular, it is desirable to propose a measure of uncertainty that similarly works for both discrete and continuous, both for one-dimensional, and for multidimensional random variables.

 *e-mail address: chraib@hotmail.com
 **e-mail address: achudnov@uic.edu

In paper of C.Tsallis [5], it was introduced the generalized entropy (Tsallis entropy), which in recent years has become very popular. This paper was not the first one, in which other entropy was proposed [6], for example Renyi entropy appeared much earlier [7], however it was Tsallis entropy that found especially many applications in physics and not only in physics [8]. There appeared non-extensive statistical mechanics [9], which Tsallis himself considered more accurately to call non-additive statistical mechanics. It should also be denoted the works on the generalization of entropy BG, devoted to superstatistics, in which the parameter corresponding to the temperature itself is considered random [10-13]. This generalization of classical statistical mechanics has also found many applications [14,15].

When using the maximum entropy principle of Jaynes [16,17], it is proposed to construct an unknown probability distribution of a random variable from the condition of the maximum of BG entropy with a few natural constraints, the issue of constraints has been studied in various papers. In papers [18,19] there are the tables that show constraints, required to obtain one or another commonly used distribution from the maximum entropy principle. For example, if we set only mean as constraint , we get an exponential distribution, if we also specify variance, we get a normal distribution. Some of the constraints required to obtain typical distributions, from our point of view, are awkward. But the most important thing is that the reason is not clear, why one should use these various constraints, if we look for different probability distributions.

We can try to do the opposite. Namely, as a constraint we use only mean, but at the same time we refuse using of BG entropy as universal. As for the exponential distribution, which as the only one can be obtained in this way by maximizing the BG entropy with mean only as constraint, it seems that BG entropy is universal for ideal gases simply because the exponential distribution for energies for ideal gases is universal. In a number of other areas in physics and not only in physics different variants of generalized entropy have been successfully used.

The structure of paper is as follows. In sec. 2 we are presented the basic ideas of the method for determining the J-measure of uncertainty in the case when the distribution of random variable is known. In sec. 3 the possibility of applying the proposed method to a two-dimensional (bivariate) random variables is discussed, in sec.4 studied the application of method to histogram. Sec. 5 contains conclusion, in Appendix shown the details of calculations for gamma distribution

## 2. J-measure of uncertainty for different probability distributions.

If the probability distribution density is given, and only the new measure of uncertainty associated with this density is required, we can use in the spirit of paper [20] a simple and direct way, namely, using the maximum entropy principle in direction, inverse to the standard, for solving this problem. Other possibility to generate a generalized entropy form is proposed in papers [21-23], where it is shown the connection between generalized entropy and the steady states solutions of the nonlinear Fokker-Planck equations.

We look for the density of J-measure of uncertainty (JMU) $g(p(x)), x)$ from the condition of the maximum of the functional $h(p)$, taking into account natural constraints such as the normalization and given mean value

$$. \qquad h(p) = \int_{P(X)} g(p)dp - \mu(\int_a^\infty p(t)dt - 1) - \nu(\int_a^\infty tp(t)dt - m), \qquad (1)$$

where $p = p(x), x \in [a, \infty], m$ - mean of probability distribution, $\mu, \nu$ -Lagrange multipliers. Since the constraints are fulfilled practically automatically for given distribution, we obtain follows

$$\frac{\partial g}{\partial p} = \mu + \nu x \qquad (2)$$

After integration over $p$ we'll obtain

$$g(p,x) = \int_0^p (\mu + \nu x)dp = \int_a^x (\mu + \nu t)p'(t)dt + C_1 = \mu p(x) + \nu \int_a^x tp'(t)dt + C_1 =$$
$$= \mu p(x) + \nu[xp(x) - F(x)] + C_1 = (\mu + \nu x)p(x) - \nu F(x) + C_1, \qquad (3)$$

where $F(x)$ - cumulative distribution function. We note that a new JMU density $g(p, x)$ is a functional of the probability density $p(x)$ and the cumulative probability distribution $F(x)$, also may depend on $x \in [a, \infty]$ directly. Integrating (3), we'll obtain JMU of system $S(x)$ as function of random variable $x$ and the full measure of uncertainty of system $S$

$$S(x) = \int_a^x g(p,x)dx = (\mu - \nu x)F(x) + 2\nu \int_a^x tp(t)dt + C_1 x + C_2 \quad ; \quad S = S(\infty) \qquad (4)$$

If it is possible to find the inverse function in an explicit form, it could be obtained the desired JMU as a functional of $p(x)$, i.e. in standard form. For a multi-valued inverse function, this must be done for each segment.

Before we consider the application of the proposed approach to various probability distributions and formulate the conditions for determining arbitrary constants, we make a remark about the concavity of the JMU.

It is clear, that BG entropy is concave for any probability distribution because of following result

$$g = -p \ln p \quad ; \quad g'(p) = -(\ln p + 1) \quad ; \quad g''(p) = -\frac{1}{p} < 0 \quad . \qquad (5)$$

In the general case

$$g'(p) = \mu + \nu x \quad ; \quad g''(p) = \nu x'(p) = \frac{\nu}{p'(x)} \qquad (6)$$

i.e., for monotonously decreasing density distributions, as exponential (in this case we assume that $a = 0$), concavity is takes place with arbitrary positive $\nu$, because $p'(x) < 0$, we'll express constants $\mu, C_1, C_2$ through $\nu$ from the following conditions $g(0) = g(\infty) = S(0) = 0$, from here we obtain $\mu = -\dfrac{\nu}{p(0)}; C_1 = \nu; C_2 = 0$ and after substitution them in (3)-(4)

$$g(x) = \nu\{[x - \frac{1}{p(0)}]p(x) + 1 - F(x)\}$$

$$S(x) = \nu\{x[1 - F(x)] - \frac{F(x)}{p(0)} + 2\int_0^x tp(t)dt\}; S(\infty) = \nu(2m - \frac{1}{p(0)})$$

(7)

For unimodal distributions as the normal , gamma, etc., where there is a maximum of probability density, i.e. mode, and so the derivative changes its sign, it is impossible, choosing the same constant $\nu$, to satisfy the required for concavity condition $g''(p) \le 0$ for all $x$, i.e. for all $p$. But this can be achieved by choosing $\nu = sign(x - x_0)\nu_1$ ; $\nu_1 > 0$ where $x_0$ -the mode of distribution. Then we obtain

$$g''(p) = sign(x - x_0)\nu_1 x'(p) = \frac{sign(x - x_0)\nu_1}{p'(x)} \le 0 \quad .$$

(8)

Now we'll obtain the general formulas for entropy for unimodal probability densities. We use formulas (3)-(4) and (8), but we consider separately the cases $x \le x_0$ (Case 1) and $x \ge x_0$ (Case 2), where $x_0$ - the mode of distribution. Then we obtain

Case 1: $x \le x_0$, $\nu = -\nu_1$ , $\nu_1 > 0$

$$g(x) = (\mu_1 - \nu_1 x)p(x) + \nu_1 F(x) + C_1$$

(9)

.

$$S(x) = \int_a^x g(t)dt = (\mu_1 + \nu_1 x)F(x) - 2\nu_1\int_a^x tp(t)dt + C_1 x + C_2$$

Case 2: $x \ge x_0$, $\nu = \nu_1$ , $\nu_1 > 0$

$$g(x) = (\mu_2 + \nu_1 x)p(x) - \nu_1 F(x) + C_3$$

(10)

$$S(x) = \int_a^x g(t)dt = (\mu_2 - \nu_1 x)F(x) + 2\nu_1\int_a^x tp(t)dt + C_3 x + C_4 ,$$

where $a$ is the lower limit of integration, for example $a = -\infty$ for normal distribution, $a = 0$ for gamma distributions.

Now, leaving the constant $\nu_1$ arbitrary, we formulate six conditions to express constants $\mu_1, \mu_2, C_1, C_2, C_3, C_4$ through $\nu_1$.

$$g(a) = 0; \quad g(x_0 - 0) = 0; \quad g(x_0 + 0) = 0; \quad g(\infty) = 0; \quad S(a) = 0; \quad S(x_0 - 0) = S(x_0 + 0) \quad (11)$$

These requirements seem natural and are an extension of similar conditions for a simple case in the classics. The selection of constants plays a decisive role and should be the subject of discussion.

We assume further that $p(a) = 0$, the general case is similar to this one. Then it could be obtained as follows:

$$\mu_1 = \frac{\nu_1}{p(x_0)}[x_0 p(x_0) - F(x_0)]; \quad \mu_2 = -\frac{\nu_1}{p(x_0)}[x_0 p(x_0) - F(x_0) + 1];$$

$$C_1 = C_2 = 0; \quad C_3 = \nu_1; \tag{12}$$

$$C_4 = \nu_1\{[4x_0 - \frac{2F(x_0)}{p(x_0)} + \frac{1}{p(x_0)}]F(x_0) - x_0 - 4\int_a^{x_0} tp(t)dt\},$$

After substitution of constants we obtain

Case 1

$$g(x) = \nu_1\{[x_0 - x - \frac{F(x_0)}{p(x_0)}]p(x) + F(x)\}$$

$$S(x) = \nu_1\{[x_0 + x - \frac{F(x_0)}{p(x_0)}]F(x) - 2\int_a^x tp(t)dt\}$$

Case 2 $\hspace{6cm}$ (13)

$$g(x) = \nu_1\{[x - x_0 + \frac{F(x_0)}{p(x_0)} - \frac{1}{p(x_0)}]p(x) - F(x) + 1\}$$

$$S(x) = \nu_1\{[-x - x_0 + \frac{F(x_0)}{p(x_0)} - \frac{1}{p(x_0)}]F(x) + 2\int_a^x tp(t)dt + x\} + C_4,$$

where $C_4$ can be found from formula (12).

Then JMU is

$$S = S(\infty) = \nu_1\{[2x_0 - \frac{F(x_0)}{p(x_0)}][2F(x_0) - 1] - \frac{1 - F(x_0)}{p(x_0)} + 2m - 4\int_a^{x_0} tp(t)dt\} \tag{14}$$

### 2.1 Exponential distribution.

For the exponential distribution, the proposed method takes the following form. Probability density function is $p(x) = \lambda e^{-\lambda x}$, the cumulative distribution function is

$F(x) = 1 - e^{-\lambda x}$, $0 \le x < \infty$, $m = \frac{1}{\lambda}$ - mean, $\sigma^2 = \frac{1}{\lambda^2}$ - variance. After substitution $p(x)$ in (3), we obtain

$$g(x) = (\mu\lambda + \nu)e^{-\lambda x} + \nu x e^{-\lambda x} - \nu + C_1. \tag{15}$$

From formula for $p(x)$ $\quad x = -\frac{1}{\lambda}(\ln p - \ln \lambda)$, then

$$g(p) = -\frac{\nu}{\lambda} p \ln p + (\mu + \frac{\nu}{\lambda} + \frac{\nu}{\lambda}\ln \lambda)p - \nu + C_1 \quad. \tag{16}$$

If we take $C_1 = \nu = \lambda$, $\mu = -1 - \ln \lambda$, we obtain the density of BG entropy. But we aim to choose constants in a way that is the same or similar for any given distributions laws. Using (7) we'll obtain

.   $$g(x) = \nu \lambda x e^{-\lambda x} \tag{17}$$

$$S(x) = \nu[\frac{1}{\lambda}(1 - e^{-\lambda x}) - x e^{-\lambda x}], \quad S = S(\infty) = \frac{\nu}{\lambda} \tag{18}$$

If we substitute $e^{-\lambda x} = \frac{p}{\lambda}; x = -\frac{1}{\lambda} \ln \frac{p}{\lambda}$, we obtain

$$g(p) = -\nu \frac{p}{\lambda} \ln \frac{p}{\lambda} . \tag{19}$$

We'll notice that at first, $m = \sigma = \frac{1}{\lambda}$ for exponential distribution, secondly, that for all arbitrariness of $\nu$, it should not be taken $\nu = \lambda$, as was done above in the classical case. The density of BG originates from a discrete case, where $0 \le p \le 1$ and it is logical therefore to require $g(p=0) = g(p=1) = 0$, but in the continuous case it is curiously, because in this example $0 \le p \le \lambda$. Note also that $\frac{p}{\lambda}$ is the normalized probability density.

## 2.2    Normal distribution

Now we consider the normal distribution with probability density

$p(x) = \frac{1}{\sigma\sqrt{2\pi}} e^{-\frac{(x-m)^2}{2\sigma^2}}$, the cumulative distribution function is $F(x) = \frac{1}{2}[1 + erf(\frac{x-m}{\sigma\sqrt{2}})]$, where $-\infty < x < \infty$, $m, \sigma^2$ - mean and variance, $erf(x) = \frac{2}{\sqrt{\pi}} \int_0^x e^{-t^2} dt$ - error function.

For a normal distribution, the mode $x_0$ and mean $m$ coincide. To obtain the JMU density $g(p)$ as a functional of the distribution density $p(x)$, we substitute in (13) $x(p) = m \pm \sigma\sqrt{-2\ln(p\sigma\sqrt{2\pi})}$, where the upper sign corresponds to the case $x > m$, and the lower sign corresponds to the case $x \le m$. The sign for $\nu$ follows (7), i.e. $\nu = sign(x-m)\nu_1$, $\nu_1 > 0$.

Then it can be obtained $g(p)$ as functional of $p(x)$ only

$$g(p) = \nu_1\{\mp\frac{1}{2}[1 - erf(\sqrt{-\ln(p\sigma\sqrt{2\pi})})] + \sigma p\sqrt{2}\sqrt{-\ln(p\sigma\sqrt{2\pi})} \pm \sigma p\sqrt{\frac{\pi}{2}}\} , \tag{20}$$

here the rule of signs is the same as above.

To find the JMU more natural using of formulas (13)-(14), and after substitution $a = -\infty, x_0 = m, p(x_0) = \frac{1}{\sigma\sqrt{2\pi}}, F(x_0) = \frac{1}{2}$ we obtain

for $x \le m$

$$g(x) = \nu_1\{(m - x - \sigma\sqrt{\frac{\pi}{2}})\frac{1}{\sigma\sqrt{2\pi}} e^{-\frac{(x-m)^2}{2\sigma^2}} + \frac{1}{2}[1 + erf(\frac{x-m}{\sigma\sqrt{2}})]\} , \tag{21}$$

$$S(x) = \nu_1\{\frac{1}{2}(x - m - \sigma\sqrt{\frac{\pi}{2}})[1 + erf(\frac{x-m}{\sigma\sqrt{2}})] + \sigma\sqrt{\frac{2}{\pi}}e^{-\frac{(x-m)^2}{2\sigma^2}}\}$$

. for $x \geq m$

$$g(x) = v_1 \{ (x - m - \sigma\sqrt{\frac{\pi}{2}}) \frac{1}{\sigma\sqrt{2\pi}} e^{-\frac{(x-m)^2}{2\sigma^2}} + \frac{1}{2}[1 - erf(\frac{x-m}{\sigma\sqrt{2}})] \} ,$$ (21a)

$$S(x) = v_1 \{ \frac{1}{2}(-x + m - \sigma\sqrt{\frac{\pi}{2}})[1 + erf(\frac{x-m}{\sigma\sqrt{2}})] + \sigma\sqrt{\frac{2}{\pi}} e^{-\frac{(x-m)^2}{2\sigma^2}} + x - m + 2\sigma\sqrt{\frac{2}{\pi}} \}$$

The result is

$$S = S(\infty) = v_1 \sigma (2\sqrt{\frac{2}{\pi}} - \sqrt{\frac{\pi}{2}}) \approx 0{,}342 v_1 \sigma$$ (22)

So for the normal distribution, the proposed JMU is proportional to the standard deviation of the random variable, it is, we must admit, quite unexpectedly.

It is well known that the classical BG entropy for a normal distribution is $S_{BG} = \frac{1}{2}\ln(2\pi e\sigma^2)$, where $\sigma^2$ - variance. Jaynes measure of uncertainty (JMU) is shown on FIG.1, for comparison with JMU (22) proposed here it is shown BG entropy too.. As expected, both grow, with the growth of standard deviation, however, in a neighborhood of zero they behave in a opposite way. BG entropy $\rightarrow -\infty$, and JMU $\rightarrow 0$, which seems more natural, because if $\sigma \rightarrow 0$ a normal density distribution $p(x) \rightarrow \delta(x - m)$, where $\delta(x)$ - Dirac delta function, i.e. the random variable becomes deterministic

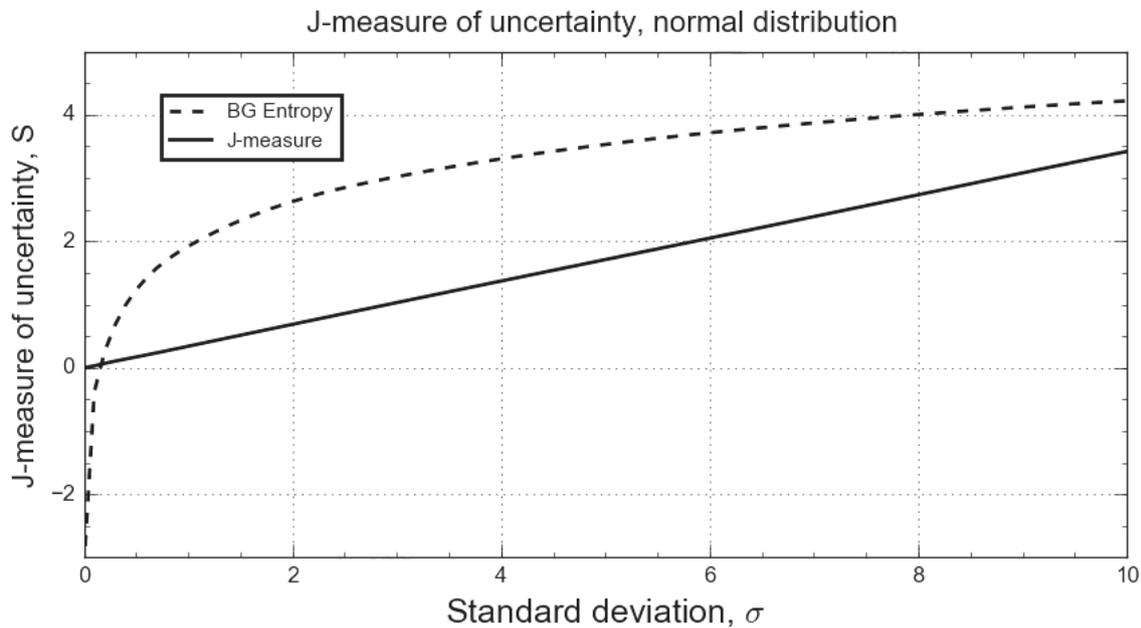

FIG.1   J-measure of normal distribution

## 2.3 Gamma distribution

The gamma distribution density and cumulative distribution function are

$$p(x) = bx^{\alpha-1}e^{-\beta x}, \quad F(x) = \frac{1}{\Gamma(\alpha)}\gamma(\alpha,\beta x) \tag{23}$$

where $\alpha, \beta$ - shape and rate parameters; mean is $m = \frac{\alpha}{\beta}$, the variance is $\sigma^2 = \frac{\alpha}{\beta^2}$;

the mode is $x_0 = \frac{\alpha-1}{\beta}$; $b = \frac{\beta^\alpha}{\Gamma(\alpha)}$ - the normalization constant, $\Gamma(\alpha) = \int_0^\infty t^{\alpha-1}e^{-t}dt$ - gamma function; $\gamma(s,x) = \int_0^x t^{s-1}e^{-t}dt$ - the lower incomplete gamma function. Detailed calculations are given in the appendix. All calculations were made for case $\alpha > 1$ as an example directly for the gamma distribution without using general formulas (13)-(14). Of course the results below can also be obtained by substituting the gamma distribution into general formulas. The JMU densities and JMU as functions of random value $x$ for two cases: $x \le x_0$ and $x \ge x_0$ take the following form

Case 1  $x \le x_0$

$$g_1(p) = -\nu_1\{\frac{1}{\Gamma(\alpha)}[(\alpha-1)\gamma(\alpha,-(\alpha-1)W) - \gamma(\alpha+1,-(\alpha-1)W)] -$$
$$- \frac{p}{\beta}(\frac{e}{\alpha-1})^{\alpha-1}[(\alpha-1)\gamma(\alpha,\alpha-1) - \gamma(\alpha+1,\alpha-1)]\} \tag{24}$$

$$g_1(x) = \frac{\nu_1}{\Gamma(\alpha)}\{[\alpha-1-\gamma(\alpha,\alpha-1)(\frac{e}{\alpha-1})^{\alpha-1} - \beta x](\beta x)^{\alpha-1}e^{-\beta x} + \gamma(\alpha,\beta x)\} \tag{25}$$

$$S_1(x) = \frac{\nu_1}{\beta\Gamma(\alpha)}\{[-(\alpha+1) - \gamma(\alpha,\alpha-1)(\frac{e}{\alpha-1})^{\alpha-1} + \beta x]\gamma(\alpha,\beta x) + 2(\beta x)^\alpha e^{-\beta x}\} \tag{26}$$

Case 2   $x \ge x_0$

$$g_2(p) = \nu_1\{-\frac{1}{\Gamma(\alpha)}[(\alpha-1)\gamma(\alpha,-(\alpha-1)W) - \gamma(\alpha+1,-(\alpha-1)W) - (\alpha-1)\Gamma(\alpha) +$$
$$+ \Gamma(\alpha+1)] - \frac{p}{\beta}(\frac{e}{\alpha-1})^{\alpha-1}[(\alpha-1)\gamma(\alpha,\alpha-1) - \gamma(\alpha+1,\alpha-1) - (\alpha-1)\Gamma(\alpha) + \Gamma(\alpha+1)]\} \tag{27}$$

$$g_2(x) = \frac{\nu_1}{\Gamma(\alpha)}\{[-(\alpha-1) + (\gamma(\alpha,\alpha-1) -$$
$$- \Gamma(\alpha))(\frac{e}{\alpha-1})^{\alpha-1} + \beta x](\beta x)^{\alpha-1}e^{-\beta x} - \gamma(\alpha,\beta x) + \Gamma(\alpha)\} \tag{28}$$

$$S_2(x) = \frac{\nu_1}{\beta \Gamma(\alpha)} \{\beta x[\Gamma(\alpha) - \gamma(\alpha, \beta x)] + [\alpha + 1 + (\gamma(\alpha, \alpha - 1) -$$

$$- \Gamma(\alpha))(\frac{e}{\alpha - 1})^{\alpha - 1}]\gamma(\alpha, \beta x) - [4 + (2\gamma(\alpha, \alpha - 1) - \quad (29)$$

$$- \Gamma(\alpha))(\frac{e}{\alpha - 1})^{\alpha - 1}]\gamma(\alpha, \alpha - 1) - 2(\beta x)^\alpha e^{-\beta x} + 4(\alpha - 1)^\alpha e^{-(\alpha - 1)} - (\alpha - 1)\Gamma(\alpha)\}$$

JMU for gamma distribution is following

$$S = S_2(\infty) = \frac{\nu}{\beta \Gamma(\alpha)} \{[\alpha + 1 + (\gamma(\alpha, \alpha - 1) - \Gamma(\alpha))(\frac{e}{\alpha - 1})^{\alpha - 1}]\Gamma(\alpha) - [4 +$$

$$+ (2\gamma(\alpha, \alpha - 1) - \Gamma(\alpha))(\frac{e}{\alpha - 1})^{\alpha - 1}]\gamma(\alpha, \alpha - 1) + 4(\alpha - 1)^\alpha e^{-(\alpha - 1)} - (\alpha - 1)\Gamma(\alpha)\}$$

(30)

In these formulas $W = -\frac{\beta x}{\alpha - 1}$ - Lambert-W function, i.e. the solving of the equation $x = W(x)e^{W(x)}$. We note, that if $\alpha > 1$, gamma distribution has maximum, if $\alpha < 1$, it is monotonically decreasing function and if $\alpha = 1$, gamma distribution turns to exponential distribution. On FIG. 2 it is shown the JMU for gamma distribution (30) for $\nu_1 = 1$, $\alpha = \beta$, i.e. $m = 1, \sigma = \frac{1}{\sqrt{\alpha}}$, and BG entropy for comparison is shown too.

For gamma distribution BG entropy is

$$S_{BG} = \alpha - \ln \beta + \ln \Gamma(\alpha) + (1 - \alpha)\psi(\alpha),$$

where $\psi(x) = \frac{\Gamma'(x)}{\Gamma(x)}$ - digamma function. As an asymptotic estimate shows, BG entropy for the gamma distribution $S_{BG} \to -\infty$ if a standard deviation $\sigma \to 0$, i.e. a normal distribution similar, at the same time, the proposed Jaynes measure of uncertainty (30) behaves plausibly, namely $S \to 0$.

Now consider the case $\alpha < 1$. In this case for all $x$ if $\nu > 0$ $g''(p) < 0$, because of $W > 0$ (see formula (A7) in Appendix), i.e. the concavity condition is satisfied. We substitute density of gamma distribution in (8), (9) and obtain

$$g(x) = \mu \frac{\beta^\alpha e^{-\beta x}}{\Gamma(\alpha)x^{1-\alpha}} + \frac{\nu}{\Gamma(\alpha)}[\beta^\alpha x^\alpha e^{-\beta x} - \gamma(\alpha, \beta x)] + C_1 \quad (31)$$

$$S(x) = \frac{1}{\Gamma(\alpha)}[(\mu - \nu x)\gamma(\alpha, \beta x) + \frac{2\nu}{\beta}\gamma(\alpha + 1, \beta x)] + C_1 x + C_2 \quad (32)$$

Unfortunately, in this case it is impossible to express uniquely, as before, constants $\mu, C_1, C_2$ through $\nu$ from the all conditions $g(0) = g(\infty) = S(0) = 0$. The reason for this is that $g(x) \to \infty$ when $x \to 0$, so for the finiteness of the entropy density, should be put $\mu = 0$. From condition $g(\infty) = 0$ follows $C_1 = \nu$ and from the condition $S(0) = 0$ follows $C_2 = 0$. Then $g(0) = C_1 = \nu$ and after substitution of constants we obtain

$$g(x) = \frac{\nu}{\Gamma(\alpha)}[\Gamma(\alpha) - \gamma(\alpha, \beta x) + \beta^\alpha x^\alpha e^{-\beta x}] \qquad (33)$$

$$S(x) = \frac{\nu}{\Gamma(\alpha)}\{[\Gamma(\alpha) - \gamma(\alpha, \beta x)]x + \frac{2}{\beta}\gamma(\alpha + 1, \beta x)\} \qquad (34)$$

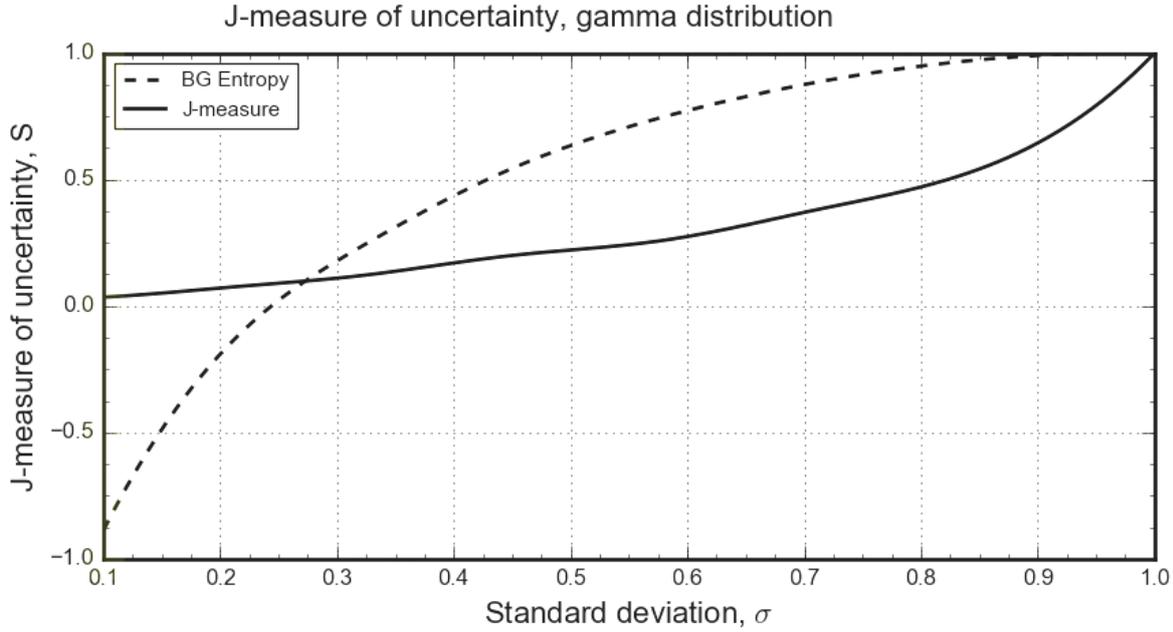

FIG.2 J-measure of uncertainty for gamma distribution.

The JMU for this case is equal

$$S = S(\infty) = \frac{2\nu\alpha}{\beta} \qquad (35)$$

It is of interest to compare the limiting entropy behavior for $\alpha \to 1$ from both sides with the entropy of the exponential distribution. For $\alpha \to 1-0$ from (35) follows $S \to \frac{2\nu}{\beta}$, for $\alpha \to 1+0$ from (30) can be obtained $S \to \frac{\nu}{\beta}$, wherein it is necessary to use the limits $\lim_{\alpha \to 1+0} \Gamma(\alpha) = 1$, $\lim_{\alpha \to 1+0} \gamma(\alpha, \alpha - 1) = 0$, $\lim_{\alpha \to 1+0}(\alpha - 1)^{\alpha - 1} = 1$. For exponential distribution from (18) follows $S \to \frac{\nu}{\beta}$ too, if we take equal the means of exponential and gamma distributions, i.e. $\lambda = \beta$ for $\alpha = 1$. So for $\alpha \to 1-0$ and for selected conditions to determine arbitrary constants JMU has a gap. However, if we'll save the requirement $C_1 = \nu$, $C_2 = 0$ but waive the requirement $\mu = 0$, we obtain from (32) for $x \to \infty$ $S = \mu + \frac{2\nu\alpha}{\beta}$, if we choose $\mu = -\frac{\nu}{\beta}$, we obtain for $\alpha \to 1-0$ $S \to \frac{\nu}{\beta}$, i.e. continuity of entropy is provided on both sides, however, $g(0)$ just like the distribution density $p(0)$ for $\alpha < 1$ is infinite.

### 2.4 Tsallis distribution

As an example of the application of the proposed method in the discrete case, we consider Tsallis distribution [26]

$$p_i = \frac{1}{Z_q}[1 - \beta(q-1)x_i]^{\frac{1}{q-1}}, \quad Z_q = \sum_{i=1}^{n}[1 - \beta(q-1)x_i]^{\frac{1}{q-1}}, \tag{36}$$

where $x_i$ - discrete random variable, $i = 1, 2, ..., n$.

To obtain the corresponding set of entropy values $g_i$ we'll look for maximum of the function $h(p_1, p_2, ..., p_W)$ with constraints

$$h(p_1, p_2, ..., p_W) = \sum_{i=1}^{n} g_i(p_i) - \mu_i(\sum_{i=1}^{n} p_i - 1) - \nu_i(\sum_{i=1}^{n} p_i x_i - m), \tag{37}$$

where m – the mean of random variable. From maximum condition

$$g'_i(p_i) = \mu_i + \nu_i x_i,$$

and after substitution $x_i = \frac{1}{\beta(q-1)}(1 - Z^{q-1}p_i^{q-1})$ from (36) we obtain

$$g'_i(p_i) = \mu_i + \frac{\nu_i}{\beta(q-1)}(1 - Z^{q-1}p_i^{q-1}). \tag{38}$$

When choosing $\frac{\nu_i}{\beta} > 0$ $g_i$ is concave since $g_i'' = -\frac{\nu_i Z^{q-1}}{\beta}p_i^{q-2} < 0$.

After integration of (38) could be obtained

$$g_i = (\mu_i + \frac{\nu_i}{\beta(q-1)})p_i - \frac{\nu_i Z^{q-1}}{\beta q(q-1)}p_i^q + C.$$

From the standard for the discrete case conditions $g_i(0) = 0, g_i(1) = 0$ follows that $C = 0$, $\mu_i = \frac{\nu_i}{\beta(q-1)}(\frac{Z^{q-1}}{q} - 1)$ and after substitution $\mu_i$ in $g_i$ we obtain

$$g_i = \frac{\nu_i Z^{q-1}}{\beta q(q-1)}(p_i - p_i^q) \tag{39}$$

Then JMU is the sum

$$S_q = \sum_{i=1}^{w} g_i = \frac{k}{q-1}(1 - \sum_{i=1}^{n} p_i^q), \quad k = \frac{\nu_i Z^{q-1}}{\beta q}, \tag{40}$$

this is exactly Tsallis entropy.

## 3. J-measure of uncertainty for two-dimensional random variables

Here we'll show the generalization of the proposed method to the multidimensional case. It seemed sufficient to consider a two-dimensional random variable, even in this simplest, but not at all simple case, a number of computational difficulties arise. After obtained general formulas we consider as examples bivariate exponential and bivariate normal distributions.

For a given bivariate distribution density of random variables, we are looking for a JMU as an integral of its density over the set of variables

$$S = \iint_{X\ Y} g(p(x,y))dxdy \qquad (41)$$

Usually, $X = (a,x), Y = (a,y)$, where $a = 0$ or $a = -\infty$, besides three constraints are given: normalization and known means of random variables $x, y$

$$\iint_{X\ Y} p(x,y)dxdy = 1 \ ; \qquad \int_X xp(x)dx = m_x \ ; \qquad \int_Y yp(y)dy = m_y \qquad (42)$$

Thus, by the method of Lagrange multipliers we are looking the maximum of the functional

$$h(p) = \int_{P(X,Y)} g(p)dp - \mu(\iint_{X\ Y} p(x,y)dxdy - 1) - \nu(\iint_{X\ Y} xp(x,y)dxdy - m_x) -$$

$$- \varepsilon(\iint_{X\ Y} yp(x,y)dxdy - m_y) \qquad (43)$$

Since the density distribution is given, we search $g(p(x,y),x,y)$ from the condition

$$h_p^{'}(p,x,y) = g^{'}(p) - \mu - \nu x - \varepsilon y = 0 \qquad (44)$$

Differentiation in respect to Lagrange multipliers leads to equations that are satisfied for the given probability density $p(x,y)$ automatically.

$$h_\mu^{'} = \iint_{X\ Y} p(x,y)dxdy - 1 = 0; \quad h_\nu^{'} = \iint_{X\ Y} xp(x,y)dxdy - m_x = 0$$

$$h_\varepsilon^{'} = \iint_{X\ Y} yp(x,y)dxdy - m_y = 0 \qquad (45)$$

We'll formulate the additional conditions for the shown in (44) and appearing later arbitrary constants. The requirement of concavity $g_p^{''} \le 0$ must be added.

As $\nu$ and $\varepsilon$ are arbitrary, it is possible to take $\varepsilon = \nu$ and introduce a new variable $z = x + y$, so the problem reduces to the case of one variable discussed above. The generalization to the case of an arbitrary number of random variables is obvious. From equation (4) we can obtain

$$g'(p) = \mu + \nu z \qquad (46)$$

Thus, in principle, the problem is solved.

Consider as the simplest example a two-dimensional exponential distribution of independent random variables, which in this case is the product of one-dimensional distributions, the density distribution and cumulative probability distribution in this case are

$$p(x, y) = \lambda_1 \lambda_2 \exp(-\lambda_1 x - \lambda_2 y) , \quad F(x, y) = [1 - \exp(-\lambda_1 x)][1 - \exp(-\lambda_2 y)] , \qquad (47)$$

it is easy to obtain the cumulative probability distribution for the sum of random variables

$$F(z) = F(x+y) = \int_0^z dx \int_0^{z-x} p(x, y) dy = 1 + \frac{\lambda_2}{\lambda_1 - \lambda_2} \exp(-\lambda_1 z) + \frac{\lambda_1}{\lambda_2 - \lambda_1} \exp(-\lambda_2 z) \qquad (48)$$

and the density distribution is

$$p(z) = F^{'}(z) = \frac{\lambda_1 \lambda_2}{\lambda_2 - \lambda_1} [\exp(-\lambda_1 z) - \exp(-\lambda_2 z)] \qquad (49)$$

Interesting is the case $\lambda_1 = \lambda_2 = \lambda$. Going to the limit $\lambda_1 \to \lambda, \lambda_2 \to \lambda$ could be obtained

$$p(z) = \lambda^2 z \exp(-\lambda z) , \quad F(z) = 1 - (1 + \lambda z) \exp(-\lambda z) \qquad (50)$$

A simple analysis shows that the density distribution (49) has a maximum at $z_0 = \frac{1}{\lambda_2 - \lambda_1} \ln \frac{\lambda_2}{\lambda_1}$ , this maximum is equal $p(z_0) = \lambda_1^{\frac{\lambda_2}{\lambda_2 - \lambda_1}} \lambda_2^{\frac{\lambda_1}{\lambda_1 - \lambda_2}}$ . Cumulative probability distribution in this point $z_0$ $F(z_0) = 1 - (\lambda_1 + \lambda_2) \lambda_1^{\frac{\lambda_1}{\lambda_2 - \lambda_1}} \lambda_2^{\frac{\lambda_2}{\lambda_1 - \lambda_2}}$ , the mean is $m_z = \frac{1}{\lambda_1} + \frac{1}{\lambda_2}$ .

Besides these functions to find of JMU by the formula (4) we need integral

$$\int_0^{z_0} z p(z) dz = -\frac{\lambda_1 \lambda_2}{\lambda_2 - \lambda_1} [\frac{1}{\lambda_1} (z_0 + \frac{1}{\lambda_1}) \exp(-\lambda_1 z_0) - \frac{1}{\lambda_2} (z_0 + \frac{1}{\lambda_2}) \exp(-\lambda_2 z_0) - \frac{1}{\lambda_1^2} + \frac{1}{\lambda_2^2}]$$

Further, the entropy is calculated by the formula (4), but we will give here as an example with a less bulky result for the particular case $\lambda_1 = \lambda_2 = \lambda$, in this case

$$z_0 = \lambda^{-1}, \quad p(z_0) = \lambda e^{-1}, \quad F(z_0) = 1 - 2e^{-1}, \quad m_z = 2\lambda^{-1}, \quad \int_0^{z_0} z p(z) dz = \lambda^{-1} (2 - \frac{5}{e}) ,$$

after substitution these values in (14), we'll obtain

$$S = \frac{V_1}{\lambda} (2 - e + 4e^{-1}) \approx 0{,}75 \frac{V_1}{\lambda} \qquad (51)$$

To analyze the effect of correlation on the entropy for bivariate exponential distribution, we choose one of the distributions proposed by Gumbel [24], we choose this distribution from many bivariate exponential distributions [25] only due to its simplicity. This density distribution is

$$p(x, y) = \exp(-x - y)\{1 + \alpha[2\exp(-x) - 1][2\exp(-y) - 1]\} , \qquad (52)$$

a cumulative distribution function

$$F(x, y) = [1 + \alpha \exp(-x - y)][1 - \exp(-x)][1 - \exp(-y)] \tag{53}$$

Means, variances and correlation coefficient respectively are

$$m_x = m_y = 1 \; ; \; \sigma_x^2 = \sigma_y^2 = 1 \; ; \; \rho = \frac{\alpha}{4} \tag{54}$$

As before we'll find $F(z)$ и $p(z)$, where $z = x + y$, and we'll obtain

$$F(z) = 1 - (z + 1) \exp(-z) + \alpha[(3 - z) \exp(-z) - (3 + 2z) \exp(-2z)] \tag{55}$$

$$p(z) = z \exp(-z) + \alpha[(z - 4) \exp(-z) + 4(z + 1) \exp(-2z)] \tag{56}$$

$$m_z = 2; \sigma_z^2 = 2 + \frac{\alpha}{2} \tag{57}$$

From condition $p'(z) = 0$ follows the equation $1 - z_0 + \alpha[5 - z - 4(1 + 2z) \exp(-z)] = 0$ to determine $z_0 = \arg \max p(z) = z_0(\alpha)$. The solution of this equation is well approximated by a fourth degree polynomial

$z_0 = 1 - 0{,}42\alpha - 0{.}175\alpha^2 + 0{.}081\alpha^3 + 0{.}096\alpha^4$. Additionally we need the integral

$$\int_0^{z_0} z p(z) dz = 2 - (z_0^2 + 2z_0 + 2) \exp(-z_0) + \alpha[(-z_0^2 + 2z_0 + 2) \exp(-z_0) \\ - 2(z_0 + 1)^2 \exp(-2z_0)] \tag{58}$$

After the substitution of this integral, as well as $z_0$, $m_z$, $F(z_0)$ и $p(z_0)$ in (14) could be obtained the relation $S(\alpha)$ for $\nu_1 = 1$, it is shown on FIG 3.
Note that the JMU in the absence of correlation, i.e. $S(0)$, as expected, coincides with the result (51) at $\lambda = 1$, $\nu_1 = 1$.

We'll do here the short remark about bivariate normal distribution with following density distribution

$$p(x, y) = \frac{1}{2\pi \sigma_1 \sigma_2 \sqrt{1 - \rho^2}} \exp\{-\frac{1}{2(1 - \rho^2)}[\frac{(x - m_x)^2}{\sigma_x^2} - \frac{2\rho(x - m_x)(y - m_y)}{\sigma_x \sigma_y} + \\ + \frac{(y - m_y)^2}{\sigma_y^2}]\} \tag{59}$$

As it is well known, the sum of two normally distributed random variables is a normally distributed random variable, with mean equal to the sum of means, the standard deviation is $\sigma_{x+y} = \sqrt{\sigma_x^2 + \sigma_y^2 + 2\rho \sigma_x \sigma_y}$, where $\rho$ is the correlation coefficient. So we can directly use the formula (22), substituting in it the appropriate values of standard deviation.

## 4. J-measure of uncertainty for a histogram.

It was shown above, that for given probability distribution and constraints Jaynes maximum entropy principle allows to obtain the family of measures of uncertainty. In order to choose one specific measure of this family, additional conditions must be formulated, for example, the proposed ones. Naturally, the proposed method for determining this J- measure of uncertainty (JMU) can be applied not only to the probability distribution, which is usually obtained as a result of the fitting of histogram, but to the histogram directly. The algorithm depends on whether the histogram monotonously decreases with increasing of a random variable values or unimodal.

Let, as a result of measurements, obtained a sample of values of a random variable as the basis of constructed histogram. For simplicity, consider a random variable

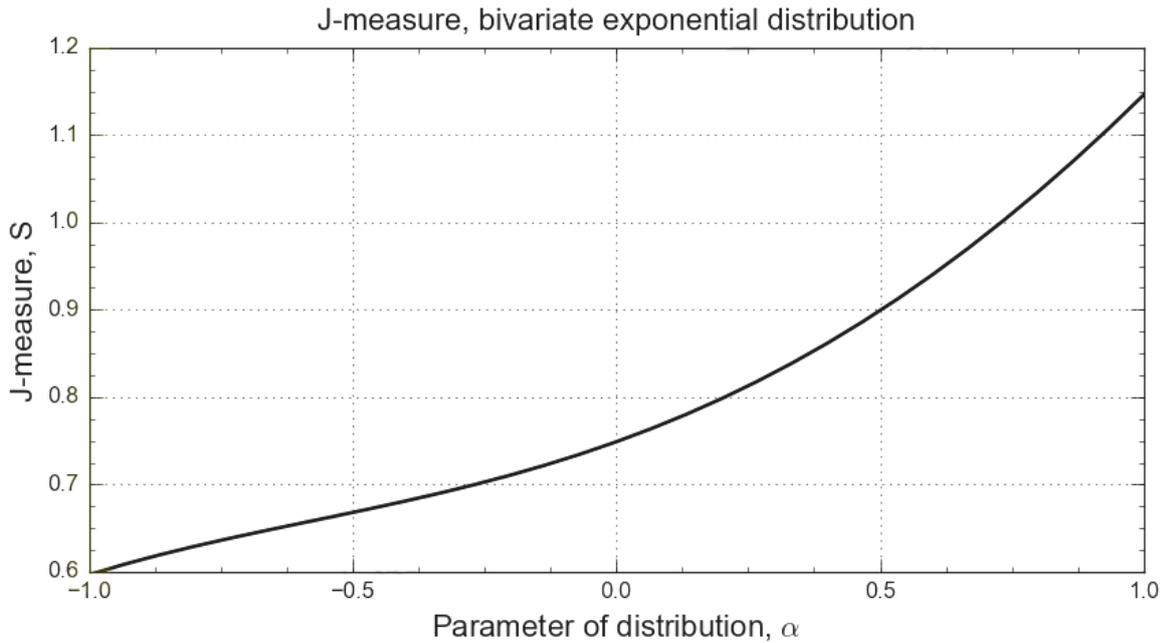

FIG 3. J-measure of bivariate exponential distribution.

distributed in interval [0,1]. The transition to the interval $[a,b]$ is obvious. The interval is divided into $n$ equal bins with length $\Delta = \dfrac{1}{n}$. Then

$x_1 = 0, x_i = (i-1)\Delta, i = 2,..,n, x_{n+1} = n\Delta = 1$; $P_1, P_2,...,P_n$, - frequencies/probabilities, $P_i = P(x_i < x < x_{i+1})$. This histogram describes the case $P_1 > P_2 > ... > P_n$. Often this type of histogram can be approximated by an exponential distribution. Another case will be studied below.

We'll transform the expression for the proposed density of JMU $g(x)$ to a form suitable for use in the case of a known histogram (see formula (3))

$$g(x) = \mu p(x) + \nu[xp(x) - F(x)] + C_1 \tag{60}$$

Density distribution $p(x)$ and cumulative distribution $F(x)$ for histogram of this type when $x_i < x < x_{i+1}$ are

$$p_i(x) = \frac{P_i}{\Delta}; F_i(x) = \sum_{k=1}^{i-1} P_k + \frac{P_i}{\Delta}(x - x_i) \tag{61}$$

After substitution (61) in (60) and transformations we'll obtain

$$g_i = \mu \frac{P_i}{\Delta} - \nu(G_i - iP_i) + C_1, \qquad (62)$$

where $G_i = \sum_{k=1}^{i} P_k$ - cumulative histogram for histogram $P_i$.

The formula (62) is an analogue of the formula (60) for histogram. As before, it is a family of JMU densities, from the concavity requirement for $g_i$, the condition $\nu = \nu_1 > 0$ must be satisfied. Up to a factor of $\nu_1$ constants $\mu$ and $C_1$ are determined from the following conditions $g_1 = g_n = 0$, then can be obtained

$$\mu = -\nu_1 \frac{\Delta - P_n}{P_1 - P_n} \quad ; \quad C_1 = \nu_1 \frac{P_1}{\Delta} \frac{\Delta - P_n}{P_1 - P_n} \qquad (63)$$

We substitute (63) into (62) and obtain .

$$g_i = \nu_1 [\frac{P_1 - P_i}{P_1 - P_n}(1 - \frac{P_n}{\Delta}) - G_i + iP_i] \qquad (64)$$

Full JMU of histogram is equal

$$S_n = \Delta \sum_{i=1}^{n} g_i = \nu_1 \Delta [\frac{1 - \frac{P_n}{\Delta}}{P_1 - P_n}(P_1 n - 1) - (n+1) + 2\sum_{i=1}^{n} iP_i] \qquad (65)$$

Consider an example. Let a histogram is given, $n = 10, \Delta = 0,1$ and the values of frequencies / probabilities are given in Table 1

Table 1. Histogram und cumulative histogram

| $i$ | 1 | 2 | 3 | 4 | 5 | 6 | 7 | 8 | 9 | 10 |
|---|---|---|---|---|---|---|---|---|---|---|
| $P_i$ | 0,150 | 0,136 | 0,123 | 0,112 | 0,101 | 0,091 | 0,083 | 0,075 | 0,068 | 0,061 |
| $G_i$ | 0,150 | 0,286 | 0,409 | 0,521 | 0,622 | 0,713 | 0,796 | 0,871 | 0,939 | 1,000 |

After substitution $n, \Delta, P_i, G_i, i = 1,...,n$ in (65) we'll obtain $S_n = 0,058\nu_1$.

Now we consider how does the approximation of a given histogram by the exponential distribution in interval [0,1] affect JMU.

For truncated exponential distribution in interval [0,1] density distribution and cumulative distribution are as follows

$$p(x) = \frac{1}{1 - e^{-\lambda}} e^{-\lambda x} \quad ; \quad F(x) = \frac{1 - e^{-\lambda x}}{1 - e^{-\lambda}} \qquad (66)$$

If we use the constants $\mu$ and $C_1$ found for histogram, JMU for this distribution can be after integration of (60) obtained

$$S = \int_0^1 g(x)dx = \nu_1 \{ \frac{1}{1 - e^{-\lambda}} [\frac{1}{\lambda}(1 - 2e^{-\lambda}) + \frac{1}{\lambda^2}(1 - e^{-\lambda}) - 1] + \frac{\Delta - P_n}{P_1 - P_n}(\frac{P_1}{\Delta} - 1) \} \qquad (67)$$

It can be shown that the truncated exponential distribution with $\lambda = 1$ approximates the histogram (Table 1) with sufficient accuracy. After substitution $\lambda = 1$, $\Delta = 0,1$, $P_1 = 0,15$, $P_n = 0,061$ in (66) we obtain $S = 0,055\nu_1$. From the comparison $S$ and $S_n$ it is clear that, despite the very accurate approximation, the JMU has decreased, as a result of the approximation, i.e. we add the additional information from the outside. Of course, it is possible to choose $\lambda$ from the condition $S = S_n$, which leads to the result $\lambda = 0,995$, however, the accuracy of approximation in this case lost. At the same time to satisfy both requirements is impossible.

It is of interest to present for comparison the results of calculating the BG entropy for the case of the histogram (Table 1) and for the case of truncated exponential distribution (66) by $\lambda = 1$

$$S_n^{BG} = -\sum_{i=1}^{n} P_i \ln P_i = 2,263 \quad ; \quad S^{BG} = -\int_{0}^{1} \frac{e^{-x}}{1 - e^{-1}} \ln(\frac{e^{-x}}{1 - e^{-1}}) dx = -0,04 \qquad (68)$$

It is easy to see that if the proposed J- measure of uncertainty for histogram and the approximating probability distribution are almost equal, although the classical BG entropies for them differ greatly, including the sign.

Now we consider the following problem. Let the results presented in Table.1 show the results of measuring any physical quantity specified in the interval [0,1]. Let, for example, the total number of measurements $m$, we denote $m_i$ the number of results that fall in the bin $i$. It is clear that they could be only integer, so Table. 1 slightly modified (Table 2). Substituting the values $n$, $\Delta$ and $P_i$ in formula (65) it can be obtained $S_{10}^{100} = 0,062\nu_1$

Table. 2. Modified histogram und cumulative histogram

| i | 1 | 2 | 3 | 4 | 5 | 6 | 7 | 8 | 9 | 10 |
|---|---|---|---|---|---|---|---|---|---|----|
| $m_i$ | 15 | 14 | 12 | 11 | 10 | 9 | 8 | 8 | 7 | 6 |
| $P_i$ | 0,15 | 0,14 | 0,12 | 0,11 | 0,10 | 0,09 | 0,08 | 0,08 | 0,07 | 0,06 |

And now we find how this value changes if we make one measurement more, i.e. now $m = 101$. To obtain this estimate, we will replace successively $m_k \rightarrow m_k' = m_k + 1$, $k = 1,...,n$ leaving the values for each of the other bins unchanged, we will find $S_{n,k}^{m'}$, and finally we will find after averaging over $k$ $S_n^{m'} = \frac{1}{n} \sum_{k=1}^{n} S_{n,k}^{m'}$. After transformations we obtain $S_{10}^{101} = 0,061\nu_1$. Thus, an additional measurement for the data of this example reduces the JMU and accordingly increases the information by about 1,6 %. In each case, before averaging over bins JMU, and respectively the information, can as well increase as decrease. It is natural to assume that the larger the initial number of measurements $m$, the less additional measurement has an impact on the entropy, and therefore on information. At the same time, when distribution of frequencies over bins remain in the same proportion, the initial JMU value does not depend on $m$, $S_{10}^{200} = S_{10}^{100}$. So for example for $m = 201$ $S_{10}^{201} = 0,0617\nu_1$, i.e. approximately 0,48%

Similarly to the previous, consider the case of a unimodal histogram, as the normal or gamma distribution. Table 3 shows the corresponding histogram, i.e. frequencies and cumulative frequencies.

Formulas for $x \le x_k$ and $x > x_k$ ( $x_k$ - the mode of distribution), similar to (9), (10) are as follows

$$g_{1i} = \mu_1 \frac{P_i}{\Delta} + \nu_1 (G_i - iP_i) + C_1 \ , \ i = 1,2,\ldots,k \tag{69}$$

$$g_{2i} = \mu_2 \frac{P_i}{\Delta} - \nu_1 (G_i - iP_i) + C_2 \ , \ i = k+1,\ldots,n$$

Table 3. Unimodal histogram and cumulative histogram.

| $i$ | 1 | 2 | 3 | 4 | 5 |
|-----|---|---|---|---|---|
| $P_i$ | $1,6 \bullet 10^{-5}$ | $8,7 \bullet 10^{-4}$ | 0,0171 | 0,1295 | 0,3521 |
| $G_i$ | $1,6 \bullet 10^{-5}$ | $8,86 \bullet 10^{-4}$ | 0,0184 | 0,1479 | 0,5 |
| $i$ | 6 | 7 | 8 | 9 | 10 |
| $P_i$ | 0,3521 | 0,1295 | 0.0171 | $8,7 \bullet 10^{-4}$ | $1,6 \bullet 10^{-5}$ |
| $G_i$ | 0,8521 | 0,9816 | 0,9951 | 0,99996 | 1 |

The constraints are $g_{11} = g_{1k} = g_{2,k+1} = g_{2n} = 0$

From here the constants can be found

$$\mu_1 = -\nu_1 \frac{\Delta(G_k - kP_k)}{P_k - P_1}, \ \mu_2 = -\nu_1 \frac{\Delta(1 - G_{k+1}) + \frac{k+1}{n}P_{k+1} - P_n}{P_{k+1} - P_n} \tag{70}$$

$$C_1 = -\mu_1 \frac{P_1}{\Delta}, \ C_2 = -\mu_2 \frac{P_n}{\Delta} + \nu_1 (1 - nP_n) \ .$$

After summing (69) over $i$ we obtain

$$S_n = \Delta(\sum_{i=1}^{k} g_{1i} + \sum_{i=k+1}^{n} g_{2i}) = \mu_1 G_k + \mu_2 (1 - G_k) + \Delta\{kC_1 +$$

$$+ (n-k)C_2 + \nu_1 [\sum_{i=1}^{k}(G_i - iP_i) - \sum_{i=k+1}^{n}(G_i - iP_i)]\} \tag{71}$$

After substitution of all necessary values in (71) can be obtained $S_n = 0,025\nu_1$.

Now compare, as before, the obtained JMU $S_n$ with the JMU found after approximation of histogram with a truncated exponential distribution in interval [0,1]

$$p(x) = A \exp[-\frac{(x-m)^2}{2\sigma^2}] \ ; \ F(x) = A\sigma \sqrt{\frac{\pi}{2}} [erf(\frac{x-m}{\sigma\sqrt{2}}) + erf(\frac{m}{\sigma\sqrt{2}})], \tag{72}$$

where $A = \{\sigma \sqrt{\frac{\pi}{2}} [erf(\frac{1-m}{\sigma\sqrt{2}})] + erf(\frac{m}{\sigma\sqrt{2}})]\}^{-1}$.. The fitting leads to the following results: $m = 0,5$, $\sigma = 0,1$, $A = 3,99$. Calculating JMU for truncated normal distribution, using the constants $\mu_1, \mu_2, C_1, C_2$, obtained for histogram, and (9), (10) leads to result $S^{norm} = 0,018\nu_1$. Thus, from the comparison it is clear that, as in the case, discussed above, the approximation uses additional information that is not contained in the original histogram. Now, as it was done for a monotone decreasing histogram, let us estimate how varies the JMU or the amount of information obtained from the additional measurement. Original number of measurements taken again $m = 100$. In Table 4 shown values of $m_i$, $P_i$ and $G_i$. To estimate the JMU from the data in this table, we as above define constants $\mu_1$, $\mu_2$, $C_1$, $C_2$ and then use the formula (71). We obtain $S_{10}^{100} = 0,0409\nu_1$. To evaluate the impact of an additional measurement, we repeat the procedure described above, i.e. for the case $m = 101$, we one after another add an additional result to each bin, in each case we calculate the JMU, then we find the average. Exactly the same as before in each case, before averaging over bins JMU and respectively the information can as well increase as decrease. As a result of this procedure we obtain $S_{10}^{101} = 0,0401\nu_1$. In this way JMU decreased, the information increased ~ 1,96 %. So on average, an additional measurement leads to the small increase of information.

Table 4 Modified unimodal histogram and cumulative histogram.

| $i$ | 1 | 2 | 3 | 4 | 5 | 6 | 7 | 8 | 9 | 10 |
|---|---|---|---|---|---|---|---|---|---|---|
| $m_i$ | 3 | 5 | 8 | 12 | 17 | 22 | 15 | 9 | 6 | 3 |
| $P_i$ | 0,03 | 0,05 | 0,08 | 0,12 | 0,17 | 0,22 | 0,15 | 0,09 | 0,06 | 0,03 |
| $G_i$ | 0,03 | 0,08 | 0,16 | 0,28 | 0,45 | 0,67 | 0,82 | 0,91 | 0,97 | 1 |

It should be noted that all the examples given in this section have only qualitative meaning and do not claim to be quantitatively accurate. They are intended to show that the proposed measure of uncertainty leads to plausible results.

## 5. Conclusion

For the case of the known distribution of a random variable from experiments or simulations, in this paper, instead of classical BG entropy, considered alternative J-measure of uncertainty (JMU).

It is obtained explicitly with the help Jaynes maximum entropy principle, used in the direction, opposite to the generally accepted. In all cases, i.e. for any probability distributions as constraints used only the normalization and the given mathematical expectation (mean). In order to use formulas (7) or (13)-(14), which make possible to find the density of the proposed JMU, it is sufficient to know the distribution of random value. An important advantage of using this measure of uncertainty in comparison with the distribution law itself is the direct information content, which is ensured by the principle of maximum entropy. On the other hand, unlike the classical Boltzmann-Gibbs entropy, based on certain assumptions, explicitly formulated or accepted by default, on the properties of a random variable, this JMU is based only on considerations that underlie the maximum entropy principle and natural additional conditions for defining arbitrary constants.

Shown the application of proposed method to exponential, normal and gamma distributions and to bivariate random variable as a example of the multidimensional case. It was made a comparison of the information contained in the histogram of a random variable with the information in the probability distribution obtained as a fitting of this histogram. Moreover, studied the influence of an additional measurement of a certain physical quantity on the amount of information.

## Appendix. J-measure of uncertainty for gamma distribution.

The gamma distribution density is

$$p(x) = bx^{\alpha-1}e^{-\beta x},$$  (A1)

where $\alpha, \beta$ - shape and rate parameters. In this appendix we consider the case $\alpha > 1$. The mean is $m = \dfrac{\alpha}{\beta}$, the variance is $\sigma^2 = \dfrac{\alpha}{\beta^2}$, the mode is $x_0 = \dfrac{\alpha-1}{\beta}$,

$b = \dfrac{\beta^\alpha}{\Gamma(\alpha)}$ - the normalization constant, $\Gamma(\alpha) = \int\limits_0^\infty t^{\alpha-1}e^{-t}dt$ - gamma function. In accordance with the maximum entropy principle, must be found the maximum of following functional

$$h(p) = \int\limits_P g(p) - \mu(\int\limits_P pdp - 1) - \nu(\int\limits_P xpdp - m),$$  (A2)

where $g(p)$ - JMU density, $\mu$, $\nu = sign(x - x_0)\nu_1$ - Lagrange multipliers, $\nu_1 > 0$. The condition of maximum of $h(p)$ is

$$g'(p) = \mu + \nu x = 0.$$  (A3)

To determine $g(p)$ it's necessary to find the inverse function $x(p)$, we consider the case $\alpha > 1$, only in this case we have unimodal distribution, we introduce new variables

$$z = -\frac{1}{\alpha-1}(\frac{p\Gamma(\alpha)}{\beta})^{\frac{1}{\alpha-1}} \;, \; W = -\frac{\beta x}{\alpha-1} \quad .$$  (A4)

Then for determination of $x(p)$ from (A1) we'll obtain the equation

$$z = We^W, \quad W = W(z(p)).$$  (A5)

The solution of (A5) is the Lambert-W function. Note that Lambert-W function is two-valued function and has two branches $W_0$ and $W_{-1}$. For $W_0$ und $z$ from (A4) $-1 \le W \le 0$, for $W_{-1}$ $-\infty < W \le -1$. This should be expected, since each value $p(x)$ takes place for two values of $x$, i.e. the inverse function is two-valued. It is easy to show that the maximum of the function $p(x)$, i.e. the mode of distribution, is at point $x = x_0 = \dfrac{\alpha - 1}{\beta}$ and equal to $p_{\mathrm{mx}} = b(\dfrac{\alpha - 1}{\beta})^{\alpha - 1} e^{-(\alpha - 1)}$. Then $z_{\min} = -\dfrac{1}{\alpha - 1}(\dfrac{p_{\mathrm{mx}} \Gamma(\alpha)}{\beta})^{\frac{1}{\alpha - 1}} = -e^{-1}$. After substitution $x = -\dfrac{\alpha - 1}{\beta} W(z(p))$ in (A3) we'll obtain

$$g'(p) = \mu - \nu \frac{\alpha - 1}{\beta} W(z(p)) \tag{A6}$$

Now we consider the question about the concavity and then continue to look for $g(p)$. After differentiating of (A6) we'll obtain

$$g''(p) = -\nu \frac{\alpha - 1}{\beta} W_p^{'}(z(p))$$

From (A4) we can find $\quad z^{'}(p) = -\dfrac{1}{(\alpha - 1)^2}(\dfrac{\Gamma(\alpha)}{\beta})^{-\frac{1}{\alpha - 1}} p^{\frac{2 - \alpha}{\alpha - 1}}$,

and from (A5) after differentiation the equation $z = We^W$ can be obtained

$$W_z^{'}(z)e^W(1 + W) = 1, \quad e^W = \frac{z}{W}, \quad \text{then} \quad W_z^{'}(z) = \frac{W}{z(1 + W)},$$

$$W_p^{'}(p(z)) = \frac{dW}{dz}\frac{dz}{dp} = \frac{dz}{dp}\frac{W}{z(1 + W)}$$

Then for $g''(p)$ we obtain

$$g''(p) = -\frac{\nu}{\beta p}\frac{W}{1 + W}. \tag{A7}$$

In order to have $g'' \le 0$ must be selected $\nu = sign(x - x_0)\nu_1 = -sign(W + 1)\nu_1$, $\nu_1 \ge 0$. And now to look for $g(p)$ we integrate (A6) over $p$ and obtain

$$g(p) = \mu p + sign(W + 1)\nu_1 \frac{\alpha - 1}{\beta}\int_0^p W(p)dp + C. \tag{A8}$$

Then

$$\int_0^p W(p)dp = \frac{\alpha - 1}{c^{\alpha - 1}}\int_0^z W(z)z^{\alpha - 2}dz = \frac{\alpha - 1}{c^{\alpha - 1}}[\int_0^{W(z)} t^{\alpha - 1}e^{(\alpha - 1)t}dt + \int_0^{W(z)} t^{\alpha}e^{(\alpha - 1)t}dt] =$$

$$= -\frac{\beta}{\Gamma(\alpha)}[\int_0^{-(\alpha - 1)W} y^{\alpha - 1}e^{-y}dy - \frac{1}{\alpha - 1}\int_0^{-(\alpha - 1)W} y^{\alpha}e^{-y}dy] = -\frac{b}{\beta^{\alpha - 1}}[\gamma(\alpha, -(\alpha - 1)W) - \frac{1}{\alpha - 1}\gamma(\alpha + 1, -(\alpha - 1)W]$$

$c = -\dfrac{\beta b^{\frac{1}{\alpha - 1}}}{\alpha - 1}$, $\quad \gamma(\alpha, x) = \int_0^x t^{\alpha - 1}e^{-t}dt$ - lower incomplete gamma function. We used the substitutions $t = W(z)$, $y = -(\alpha - 1)t$. Then the required functional takes the form

$$g(p) = \mu p - \frac{sign(W+1)\nu_1}{\Gamma(\alpha)}[(\alpha-1)\gamma(\alpha, -(\alpha-1)W) - \gamma(\alpha+1, -(\alpha-1)W] + C. \qquad (A9)$$

This functional (A9) is JMU density appropriate to gamma distribution. We can substitute $p(x)$ from (A1) and $x = -\frac{\alpha-1}{\beta}W(z)$ from (A4) in (A9) in order to obtain the density of proposed J-measure as the function of the random value $x$

$$g(x) = \mu b x^{\alpha-1}e^{-\beta x} + \frac{sign(x-x_0)\nu_1}{\Gamma(\alpha)}[(\alpha-1)\gamma(\alpha, \beta x) - \gamma(\alpha+1, \beta x)] + C. \qquad (A10)$$

JMU $S(x)$ can be obtained by integrating of the JMU density $g(x)$, then we'll find the total J- measure of uncertainty as the limit $S(\infty) = S(x)$. However, before determining $S(x)$ and $S$, it is necessary to determine the constants that are still arbitrary separately for two cases: $x \leq x_0$ i.e. $-1 \leq W \leq 0$ and $x > x_0$ i.e. $W \leq -1$.

Case 1: $x \leq x_0$, $-1 \leq W \leq 0$

From the conditions: $g = 0$ at $x = 0$ or, what the same, at $p = 0$ and at $x = x_0 = \frac{\alpha-1}{\beta}$

or, what the same, $p = \frac{\beta}{\Gamma(\alpha)}(\frac{\alpha-1}{e})^{\alpha-1}$, can be obtained $C = C_1 = 0$ and

$\mu = \mu_1 = \frac{\nu_1}{\beta}[\alpha - 1 - \gamma(\alpha, \alpha-1)(\frac{e}{\alpha-1})^{\alpha-1}]$.

Substituting these constants in (A9) and (A10), we obtain

$$g_1(p) = -\nu_1\{\frac{1}{\Gamma(\alpha)}[(\alpha-1)\gamma(\alpha, -(\alpha-1)W) - \gamma(\alpha+1, -(\alpha-1)W)] -$$
$$- \frac{p}{\beta}(\frac{e}{\alpha-1})^{\alpha-1}[(\alpha-1)\gamma(\alpha, \alpha-1) - \gamma(\alpha+1, \alpha-1)]\} \qquad (A11)$$

and after the transformations could be obtained

$$g_1(x) = \frac{\nu_1}{\Gamma(\alpha)}\{[\alpha-1 - \gamma(\alpha, \alpha-1)(\frac{e}{\alpha-1})^{\alpha-1} - \beta x](\beta x)^{\alpha-1}e^{-\beta x} + \gamma(\alpha, \beta x)\} \qquad (A12)$$

After integration we'll obtain

$$S_1(x) = \frac{\nu_1}{\beta\Gamma(\alpha)}\{[-(\alpha+1) - \gamma(\alpha, \alpha-1)(\frac{e}{\alpha-1})^{\alpha-1} + \beta x]\gamma(\alpha, \beta x) + 2(\beta x)^\alpha e^{-\beta x}\} \qquad (A13)$$

Case 2: $x > x_0$, $W \leq -1$

From the conditions: $g = 0$ at $x \to \infty$ or, what the same, at $p = 0$ and at $x = x_0 = \dfrac{\alpha - 1}{\beta}$ or, what the same, at $p = \dfrac{\beta}{\Gamma(\alpha)}(\dfrac{\alpha - 1}{e})^{\alpha - 1}$ , can be obtained

$C = C_2 = \nu_1$ and $\mu = \mu_2 = \dfrac{\nu_1}{\beta}\{-(\alpha - 1) + [\gamma(\alpha, \alpha - 1) - \Gamma(\alpha)](\dfrac{e}{\alpha - 1})^{\alpha - 1}\}$ .

Substituting these constants in (A9) and (A10), we obtain

$$g_2(p) = \nu_1\{-\frac{1}{\Gamma(\alpha)}[(\alpha - 1)\gamma(\alpha, -(\alpha - 1)W) - \gamma(\alpha + 1, -(\alpha - 1)W) - (\alpha - 1)\Gamma(\alpha) + \Gamma(\alpha + 1)] -$$

$$-\frac{p}{\beta}(\frac{e}{\alpha - 1})^{\alpha - 1}[(\alpha - 1)\gamma(\alpha, \alpha - 1) - \gamma(\alpha + 1, \alpha - 1) - (\alpha - 1)\Gamma(\alpha) + \Gamma(\alpha + 1)]\}$$

(A14)

and after the transformations could be obtained

$$g_2(x) = \frac{\nu_1}{\Gamma(\alpha)}\{[-(\alpha - 1) + (\gamma(\alpha, \alpha - 1) -$$

$$-\Gamma(\alpha))(\frac{e}{\alpha - 1})^{\alpha - 1} + \beta x](\beta x)^{\alpha - 1}e^{-\beta x} - \gamma(\alpha, \beta x) + \Gamma(\alpha)\}$$

(A15)

After integration we'll obtain

$$S_2(x) = \frac{\nu_1}{\beta\Gamma(\alpha)}\{\beta x[\Gamma(\alpha) - \gamma(\alpha, \beta x)] + [\alpha + 1 + (\gamma(\alpha, \alpha - 1) - \Gamma(\alpha))(\frac{e}{\alpha - 1})^{\alpha - 1}]\gamma(\alpha, \beta x) -$$

$$-[4 + (2\gamma(\alpha, \alpha - 1) - \Gamma(\alpha))(\frac{e}{\alpha - 1})^{\alpha - 1}]\gamma(\alpha, \alpha - 1) - 2(\beta x)^{\alpha}e^{-\beta x} +$$

$$+4(\alpha - 1)^{\alpha}e^{-(\alpha - 1)} - (\alpha - 1)\Gamma(\alpha)\}$$

(A16)

and the J-measure of system is

$$S = S_2(\infty) = \frac{\nu}{\beta\Gamma(\alpha)}\{[\alpha + 1 + (\gamma(\alpha, \alpha - 1) - \Gamma(\alpha))(\frac{e}{\alpha - 1})^{\alpha - 1}]\Gamma(\alpha) - [4 +$$

$$+(2\gamma(\alpha, \alpha - 1) - \Gamma(\alpha))(\frac{e}{\alpha - 1})^{\alpha - 1}]\gamma(\alpha, \alpha - 1) + 4(\alpha - 1)^{\alpha}e^{-(\alpha - 1)} - (\alpha - 1)\Gamma(\alpha)\}$$

(A17)